# Public Perception of Climate Change and the New Climate Dice

**James Hansen**[a1], **Makiko Sato**[a], **Reto Ruedy**[b]

[a]NASA Goddard Institute for Space Studies and Columbia University Earth Institute, [b]Trinnovim LLC, New York, NY 10025

Summary. Should the public be able to recognize that climate is changing, despite the notorious variability of weather and climate from day to day and year to year? We investigate how the probability of unusually warm seasons has changed in recent decades, with emphasis on summer, when changes are likely to have the greatest practical effects. We show that the odds of an unusually warm season have increased greatly over the past three decades, but also the shape of the frequency distribution has changed so as to enhance the likelihood of extreme events. A new category of hot summertime outliers, more than three standard deviations (3σ) warmer than climatology, has emerged, with the occurrence of these outliers having increased 1-2 orders of magnitude in the past three decades. Thus we can state with a high degree of confidence that extreme summers, such as those in Texas and Oklahoma in 2011 and Moscow in 2010, are a consequence of global warming, because global warming has dramatically increased their likelihood of occurrence.

We illustrate observed variability of seasonal mean surface air temperature anomalies in units of standard deviations, including comparison with the normal distribution ("bell curve") that the lay public may appreciate. We take 1951-1980 as an appropriate base period, because temperatures then were within the Holocene range to which humanity and other planetary life are adapted. In contrast, we infer that global temperature is now above the Holocene range, as evidenced by the fact that the ice sheets in both hemispheres are shedding mass (1) and sea level is rising at a rate [more than 3 mm/year or 3 m/millennium (2)] that is much higher than the rate of sea level change during the past several millennia.

The frequency of occurrence of local summer-mean temperature anomalies was close to the normal distribution in the 1950s, 1960s and 1970s in both hemispheres (Fig. P1A, B). However, in each subsequent decade the distribution shifted toward more positive anomalies, with the positive tail (hot outliers) of the distribution shifting more than the negative tail. The temporal change of the anomaly distribution for the contiguous United States (Fig. P1C) is similar to the global change, but much noisier because the contiguous U.S. covers only ~1.5% of the globe.

Winter warming exceeds that in summer, but the standard deviation of seasonal mean temperature at middle and high latitudes is much larger in winter (typically 2-4°C) than in summer (typically ~1°C). Thus the shift of the anomaly distribution, in the unit of standard deviations, is less in winter than in summer (Fig. P1D).

A concept of "climate dice" was suggested (3) to describe the stochastic variability of local seasonal mean temperature, with the implication that the public should recognize the existence of global warming once the dice become sufficiently "loaded" (biased). Specifically, the 10 warmest summers (Jun-Jul-Aug in the Northern Hemisphere) in the 30-year period of climatology (1951-1980) define the "hot" category, the 10 coolest the "cold" category, and the middle 10 the "average" summer. Thus it was imagined that two sides of a six-sided die were colored red, blue and white for these respective categories. The divisions between "hot" and "average" and between "average" and "cold" occur at $+0.43\sigma$ and $-0.43\sigma$ for a normal distribution of variability.

---

[1] To whom correspondence should be addressed. E-mail: james.e.hansen@nasa.gov

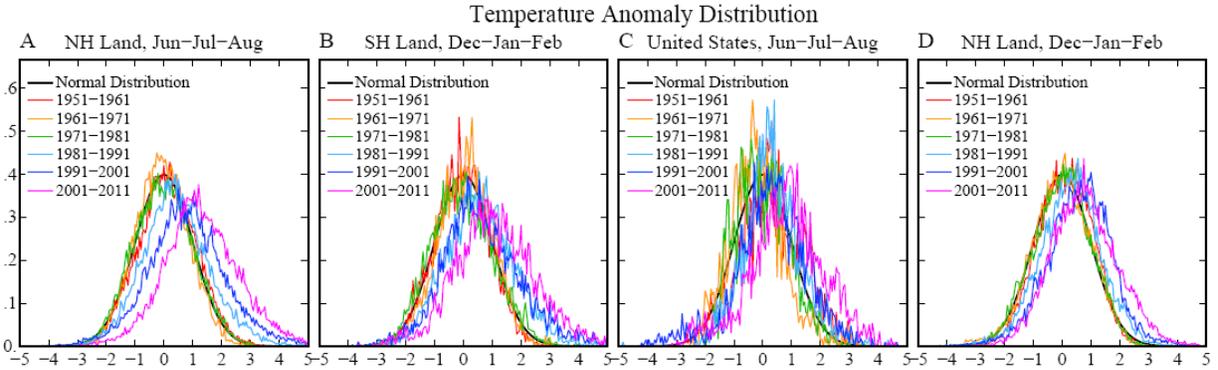

**Fig. P1.** Frequency of occurrence (y-axis) of local temperature anomalies divided by local standard deviation (x-axis) obtained by binning all local results for the indicated region and 11-year period into 0.05 frequency intervals. Area under each curve is unity. Anomalies are relative to 1951-1980 climatology, and standard deviations are for the same 1951-1980 base period.

Temperatures simulated in a global climate model reached a level such that four of the six sides of the climate dice were red in the first decade of the 21$^{st}$ century for greenhouse gas scenario B (3), which is an accurate approximation of actual greenhouse gas growth [(4), updates at http://www.columbia.edu/~mhs119/GHG_Forcing/]. We find that actual summer-mean temperature anomalies over global land during the past decade averaged about 75% in the "hot category", thus midway between four and five sides of the die were red, which is reasonably consistent with expectations.

A more important change is the emergence of a subset of the hot category, extremely hot outliers, defined as anomalies exceeding $+3\sigma$. The frequency of these extreme anomalies is about 0.13% in the normal distribution, and thus a typical summer in the period of climatology would have only about 0.1-0.2% of the globe covered by such hot extremes. We show that during the past several years the portion of global land area covered by summer temperature anomalies exceeding $+3\sigma$ has averaged about 10%, thus an increase by about a factor of 50 compared to the period of climatology. Recent examples of summer temperature anomalies exceeding $+3\sigma$ include the heat wave and drought in Oklahoma, Texas and Mexico in 2011 and a larger region encompassing much of the Middle East, Western Asia and Eastern Europe, including Moscow, in 2010.

The question of whether these extreme hot anomalies are a consequence of global warming is commonly answered in the negative, with an alternative interpretation based on meteorological patterns. For example, an unusual atmospheric "blocking" situation resulted in a long-lived high pressure anomaly in the Moscow region in 2010, and a strong La Nina in 2011 may have contributed to the heat and drought situation in the southern United States and Mexico. However, such meteorological patterns are not new and thus as an "explanation" fail to account for the huge increase in the area covered by extreme positive temperature anomalies. Specific meteorological patterns help explain *where* the high pressure regions that favor high temperature and drought conditions occur in a given summer, but the unusually great temperature extremities and the large area covered by these hot anomalies is a consequence of global warming.

This attribution is important, because we can project with a high degree of confidence that the area covered by extremely hot anomalies will continue to increase during the next few decades and even greater extremes will occur. The decade-by-decade shift to the right of the temperature anomaly frequency distribution (Fig. P1) will continue, because Earth is out of energy balance, more solar energy absorbed than heat radiation emitted to space (5), and it is this imbalance that drives the planet to higher temperatures. Even an extremely optimistic scenario, with fossil fuel emission reductions of 6%/year beginning in 2013, results in global temperature rising to almost 1.2°C relative to 1880-1920, which compares to a current level ~0.8°C.

We argue that it is important to keep the base period defining climatology fixed. Shifting the base period continually to the most recent three decades hides the increasing variability that we found. A base period prior to 1980 avoids this problem and yields a climatology within the global temperature range of the Holocene, to which nature and human civilization are adapted.

Practical effects of the increasingly loaded climate dice are likely to occur via amplification of extremes of the water cycle. Higher temperatures exacerbate hot dry conditions, but higher temperatures also increase the amount of water vapor that the atmosphere can hold. Increased water vapor leads to heavier rainfall and floods as well as the potential for stronger storms driven by latent heat including thunderstorms, tornadoes and tropical storms. We cite data suggesting that such climate impacts are already underway, but because of the small spatial scale of many of these phenomena it is necessary to gather more extensive homogeneous hydrologic data to assess ongoing global change. Such assessment is important because of potential effects on humans and other species, as it has been estimated that continued business-as-usual fossil fuel emissions and global warming could result by the end of the century in 21-52% of the species on Earth being committed to extinction IPCC (6).

**References**

1. Rignot, E., Velicogna, I., van den Broeke, M.R., Monaghan, A., and Lenaerts, J., 2011: Acceleration of the contribution of the Greenland and Antarctic ice sheets to sea level rise. *Geophys Res Lett,* **38** L05503.
2. Nerem, R.S., Leuliette, E., and Cazenave, A., 2006: Present-day sea-level change: A review. *Cr Geosci,* **338**, 1077-1083.
3. Hansen, J*., et al.*, 1988: Global climate changes as forecast by Goddard Institute for Space Studies 3-dimensional model. *J Geophys Res - Atmos,* **93**, 9341-9364.
4. Hansen, J. and Sato, M., 2004: Greenhouse gas growth rates. *Proc Natl Acad Sci USA,* **101**, 16109-16114.
5. Hansen, J., Sato, M., Kharecha, P., and von Schuckmann, K., 2011: Earth's Energy Imbalance and Implications. *Atmos Chem Phys,* **11**, 1-29.
6. Intergovernmental Panel on Climate Change (IPCC), 2007: *Climate Change 2007, Impacts, Adaptation and Vulnerability*, Parry, M.L., Canziani, O.F., Palutikof, J.P., Van Der Linden, P.J., and Hanson, C.E. eds., Cambridge Univ Press, 996 pp.

**Abstract. "Climate dice", describing the chance of unusually warm or cool seasons relative to climatology, have become progressively "loaded" in the past 30 years, coincident with rapid global warming. The distribution of seasonal mean temperature anomalies has shifted toward higher temperatures and the range of anomalies has increased. An important change is the emergence of a category of summertime extremely hot outliers, more than three standard deviations (3σ) warmer than climatology. This hot extreme, which covered much less than 1% of Earth's surface in the period of climatology, now typically covers about 10% of the land area. It follows that we can state, with a high degree of confidence, that extreme anomalies such as those in Texas and Oklahoma in 2011 and Moscow in 2010 were a consequence of global warming, because their likelihood in the absence of global warming was exceedingly small. We discuss practical implications of this substantial, growing, climate change.**

The greatest barrier to public recognition of human-made climate change is probably the natural variability of climate. How can a person discern long-term climate change, given the notorious variability of local weather and climate from day to day and year to year?

This question assumes great practical importance, because of the need for the public to appreciate the significance of human-made global warming. Actions to stem emissions of the gases that cause global warming are unlikely to approach what is needed until the public recognizes that human-made climate change is underway and perceives that it will have unacceptable consequences if effective actions are not taken to slow the climate change. A recent survey in the United States (1) confirms that public opinion about the existence and importance of global warming depends strongly on their perceptions of recent local climate variations. Early public recognition of climate change is critical. Stabilizing climate with conditions resembling those of the Holocene, the world in which civilization developed, can only be achieved if rapid reduction of fossil fuel emissions begins soon (2).

It was suggested decades ago (3) that by the early 21$^{st}$ century the informed public should be able to recognize that the frequency of unusually warm seasons had increased, because the "climate dice," describing the probability of unusually warm or unusually cool seasons, would be sufficiently loaded (biased) as to be discernible to the public. Recent high profile heat waves, such as the one in Texas and Oklahoma in the summer of 2011, raise the question of whether these extreme events are related to the on-going global warming trend, which has been attributed with a high degree of confidence to human-made greenhouse gases (4).

Summer, when most biological productivity occurs, is the most important season for humanity and thus the season when climate change may have its biggest impact. Global warming causes spring warmth to come earlier and it causes cooler conditions that initiate fall to be delayed. Thus global warming not only increases summer warmth, it also protracts summer-like conditions, stealing from both spring and fall. Our study therefore places emphasis on study of how summer temperature anomalies have been changing.

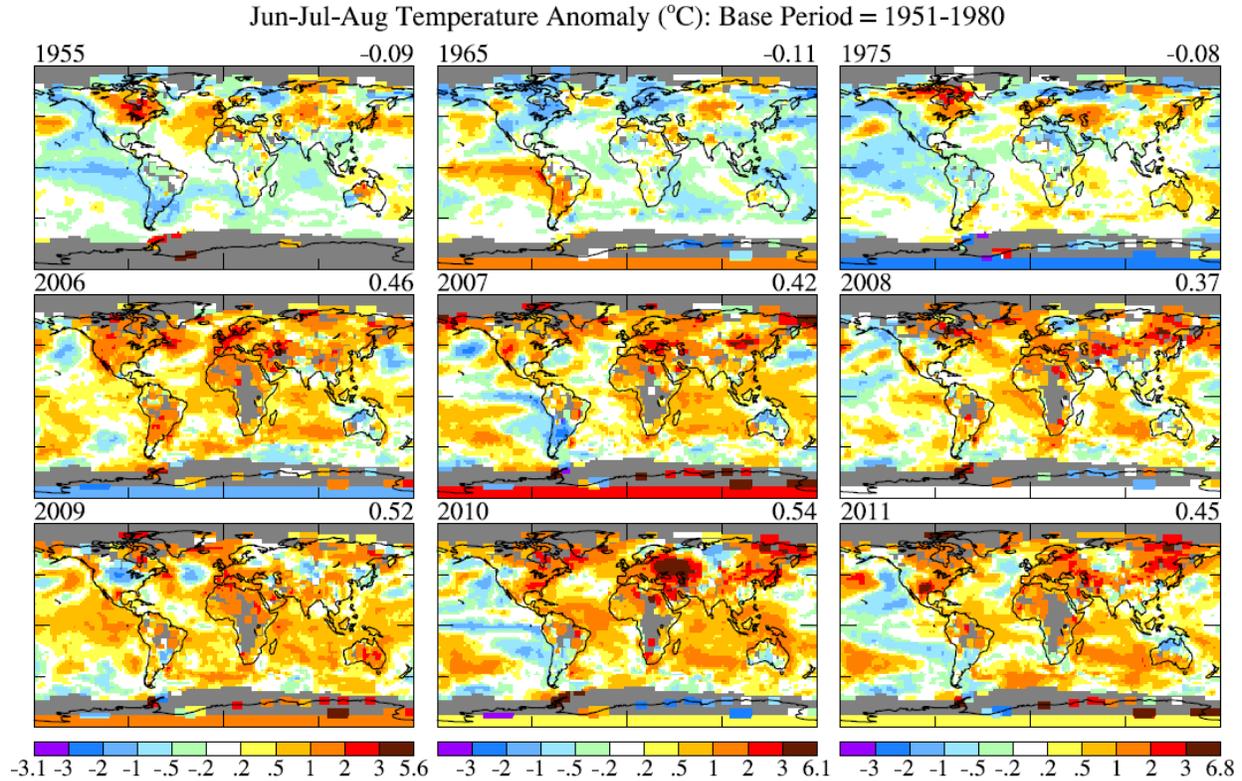

**Fig. 1.** Jun-Jul-Aug surface temperature anomalies in 1955, 1965, 1975 and the past six years relative to the 1951-1980 mean. Number on upper right is the global mean (average over all area with data).

**Materials and Methods**

We use the Goddard Institute for Space Studies (GISS) surface air temperature analysis (5) to examine seasonal mean temperature variability and how that variability has changed in recent decades. The GISS analysis is carried out at two spatial resolutions: 1200 km and 250 km. We use the 250 km analysis, because it is better-suited for illustrating seasonal-mean variability on regional spatial scales.

One of the observational records employed in the GISS analysis is the Global Historical Climatology Network (GHCN) data set for surface air temperature at meteorological stations, which is maintained by the National Oceanic and Atmospheric Administration (NOAA) National Climatic Data Center (NCDC). We use version 2 (GHCNv2) of this data record (6) here, because it is the version employed in the documented GISS analysis (5). The data record that NCDC currently provides, GHCNv3, initiated in 2011, yields a slightly larger global warming trend (0.75°C for 1900-2010, while GHCNv2 yields 0.72°C), but the changes are too small to affect the conclusions of our present study.

We illustrate observed variability of seasonal mean surface air temperature emphasizing the distribution of anomalies in units of standard deviations with comparisons to the normal distribution ("bell curve"), which the lay public may appreciate. We choose 1951-1980 as the base period for most of our illustrations, because that is a time of little global temperature trend just prior to the rapid global warming in recent decades. It is also a period that older people today, particularly those of the "baby boom" generation, can remember. Global temperature in 1951-1980 is also within the Holocene temperature range, and thus it is a climate that the natural world and civilization is adapted to. In contrast, global temperature in the first decade of the 21$^{st}$ century may already be outside the Holocene range (7), as evidenced by the fact that the Greenland and Antarctic ice sheets simultaneously are losing mass rapidly (8, 9) and sea level is now rising at a rate [3 m/millennium, (10); updates available at http://sealevel.colorado.edu/] well above the average rate during the past several thousand years.

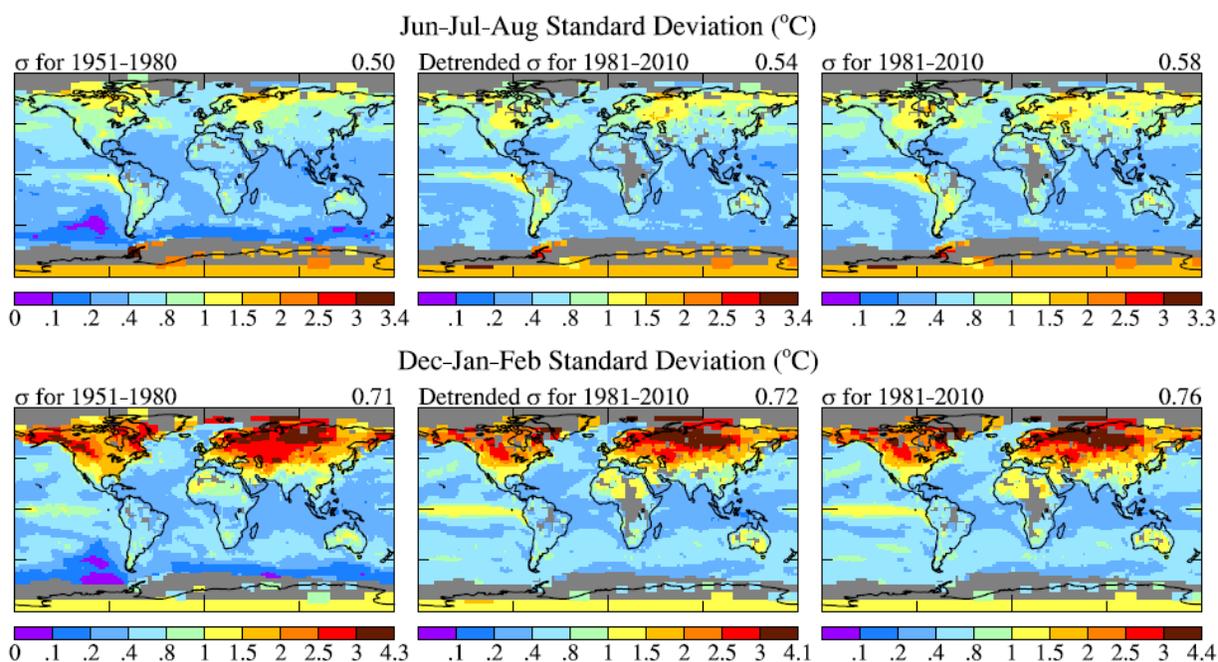

**Fig. 2.** Standard deviation of local Jun-Jul-Aug (above) and Dec-Jan-Feb (below) mean surface temperature for 30-year periods 1951-1980 (left maps) and 1981-2010. In the middle maps the local 30-year (1981-2010) temperature trend was removed before calculating the standard deviation.

**Results**

***Summer temperature anomalies.*** Jun-Jul-Aug (Northern Hemisphere summer) surface temperature anomalies relative to the base period 1951-1980 are shown in Fig. 1 for mid-decade years of the 1950s, 1960s and 1970s, and for the past six years. Most regions were warmer in recent years than during 1951-1980, but many regions still exist that are cooler than the 1951-1980 mean. The United States, for example, was unusually cool in 2009.

    What is the practical importance of such temperature anomalies? Global warming since 1951-1980 is about 0.5-0.6°C (about 1°F) (5, 11-13). This seems small, and indeed it is small compared with weather fluctuations. Yet we will suggest that this level of average warming is already having important effects.

***Natural climate variability and the standard deviation.*** A good way to gain appreciation of the warming's significance is to compare it to natural year-to-year variability of temperature. The standard deviation of local seasonal mean surface temperature over a period of years is a measure of the typical variability of the seasonal mean temperature over that period of years. Fig. 2 (left column) shows this variability during the base period 1951-1980.

    Below we will illustrate the distribution of observed temperature anomalies about their mean value. It is commonly assumed that this variability can be approximated as a normal (Gaussian) distribution, the so-called "bell curve". A normal distribution of variability has 68 percent of the anomalies falling within one standard deviation of the mean value. The tails of the normal

distribution (which we illustrate below) decrease quite rapidly so there is only a 2.3% chance of the temperature exceeding +2σ, where σ is the standard deviation, and a 2.3% chance of being colder than -2σ. The chance of exceeding +3σ is only 0.13% for a normal distribution of variability, with the same chance of a negative anomaly exceeding -3σ.

Interannual variability of surface temperature is larger in the winter hemisphere than in the summer and larger over land than over ocean (Fig. 2). The basic reason for the large winter variability is the great difference of temperature between low latitudes and high latitudes in winter. This allows the temperature at a given place to vary by tens of degrees depending on whether the wind is from the south or north. The latitudinal temperature gradient in summer is much smaller, thus providing less drive for exchange of air masses between middle latitudes and polar regions -- and when exchange occurs the effect on temperature is less than that caused by a winter "polar express" of Arctic (or Antarctic) air delivered to middle latitudes.

Note in Fig. 2 that there are areas in the Southern ocean in which the standard deviation is less than 0.1°C in both Dec-Jan-Feb and Jun-Jul-Aug. This unrealistically small variability is the result of an absence of measurements in the pre-satellite era in a region with very little ship traffic. This artifact does not occur in the standard deviation for 1981-2010 (right column in Fig. 2), when satellite observations provided uniform daily observations.

A potential drawback of using 1981-2010 to define natural variability is the existence of rapid global warming during that period, a trend that is primarily a human-made effect (4). Subtracting the local linear temperature trend before calculating the standard deviation only moderately reduces the local variability (middle column in Fig. 2). This comparison confirms that local year-to-year temperature fluctuations, not the long-term temperature trend, provide the main contribution to σ.

The global mean of the local standard deviation of Jun-Jul-Aug surface temperature increases from 0.50°C for 1951-1980 data to 0.58°C for 1981-2010 data. Only half of this increase is removed if the 1981-2010 data is detrended using the local summertime trend before the standard deviation is calculated. Indeed, the maps in Fig. 2 suggest that there are regions in the Northern Hemisphere summer where the variability is greater in 1981-2010 than in 1951-1980, even if the 1981-2010 data are detrended. The increase of variability is widespread, being apparent in North America and Asia, but also in the equatorial Pacific Ocean (Fig. 2), where the unusually strong El Ninos in 1983 and 1997-98 might be a factor.

Over the ocean, some of the increased variability could be a consequence of increased spatial and temporal resolution, because the 1981-2010 period has high-resolution satellite data, while the 1951-1980 period is largely dependent on ship data. This issue could be examined by comparing analyses based on full resolution satellite-era data with an analysis of the same period employing sub-sampling at the resolution of the pre-satellite era. However, we do not carry out such a study, because our interest is primarily in the areas where most people live. Thus in the following analyses we will focus on land data, while including some global data for comparison.

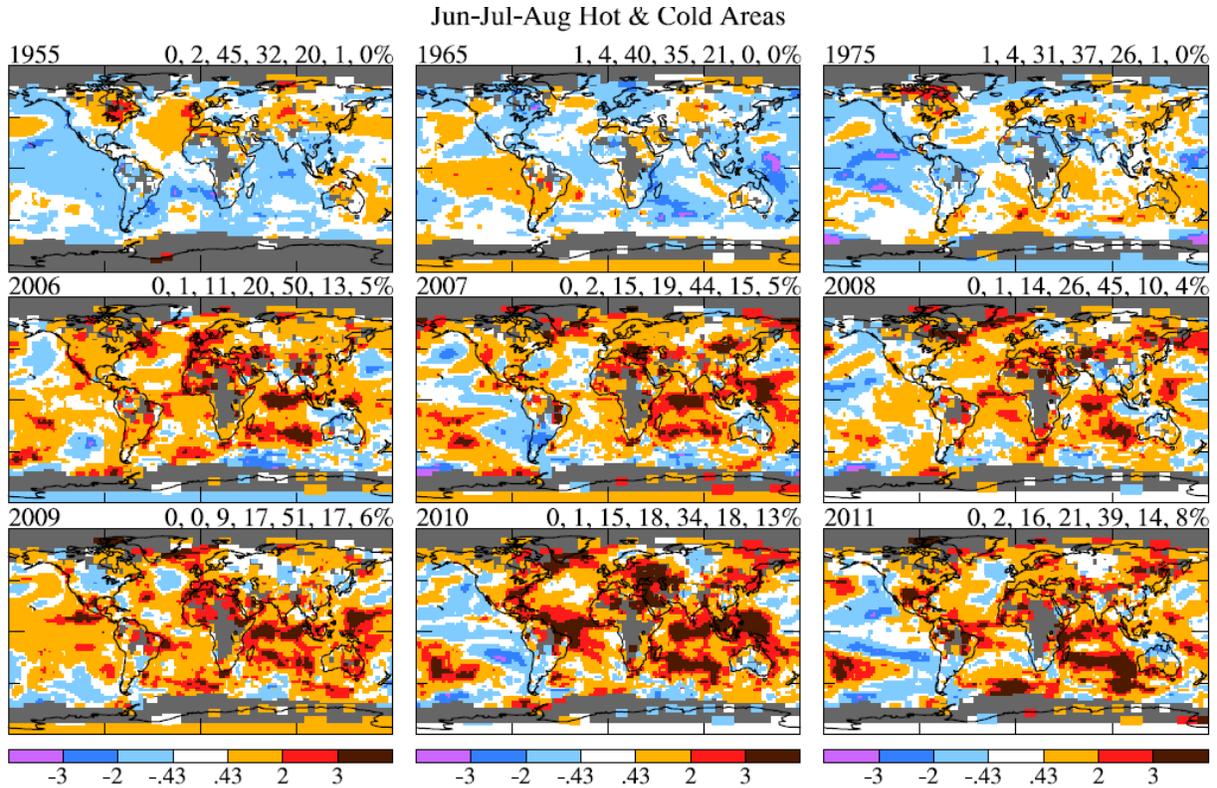

**Fig. 3.** Jun-Jul-Aug surface temperature anomalies in 1955, 1965, 1975 and in 2006-2011 relative to 1951-1980 mean temperature in units of the local detrended 1981-2010 standard deviation of temperature. Numbers above each map are percent of surface area covered by each category in the color bar.

***Recent temperature anomalies.*** Now let's examine the question: how unusual were recent Jun-Jul-Aug temperature anomalies? Fig. 3 shows the ratio: local temperature anomaly divided by local standard deviation, σ, where σ is from the middle column in Fig. 2. The red and brown areas in Fig. 3 have anomalies that exceed 2σ and 3σ, respectively. The numbers on the top of each map are the percentage of the total area covered by each of the seven categories in the color bar.

    A remarkable feature of Fig. 3 is the large brown area (anomalies > 3σ), which covered between 4% and 13% of the world in the six years 2006-2011. In the absence of climate change, and if temperature anomalies were normally distributed, we would expect the brown area to cover only 0.1-0.2% of the planet. The upper row in Fig. 3, the temperature anomalies in a mid-year of each of the three decades in the period of climatology, confirms that such extreme anomalies were practically absent in that period. Occurrence of extreme anomalies (> +3σ) in recent years is more than an order of magnitude greater than during the period of climatology, 1951-1980.

    The recent spate of 3σ events raises several questions. What does the temperature anomaly distribution look like, how is it changing, and how important is a +3σ anomaly? Well-publicized extreme conditions in Texas in 2011 and around Moscow and in the Middle East in 2010 had summer temperature anomalies reaching the +3σ level (Fig. 3), suggesting that increase of such extreme events may have large practical impacts.

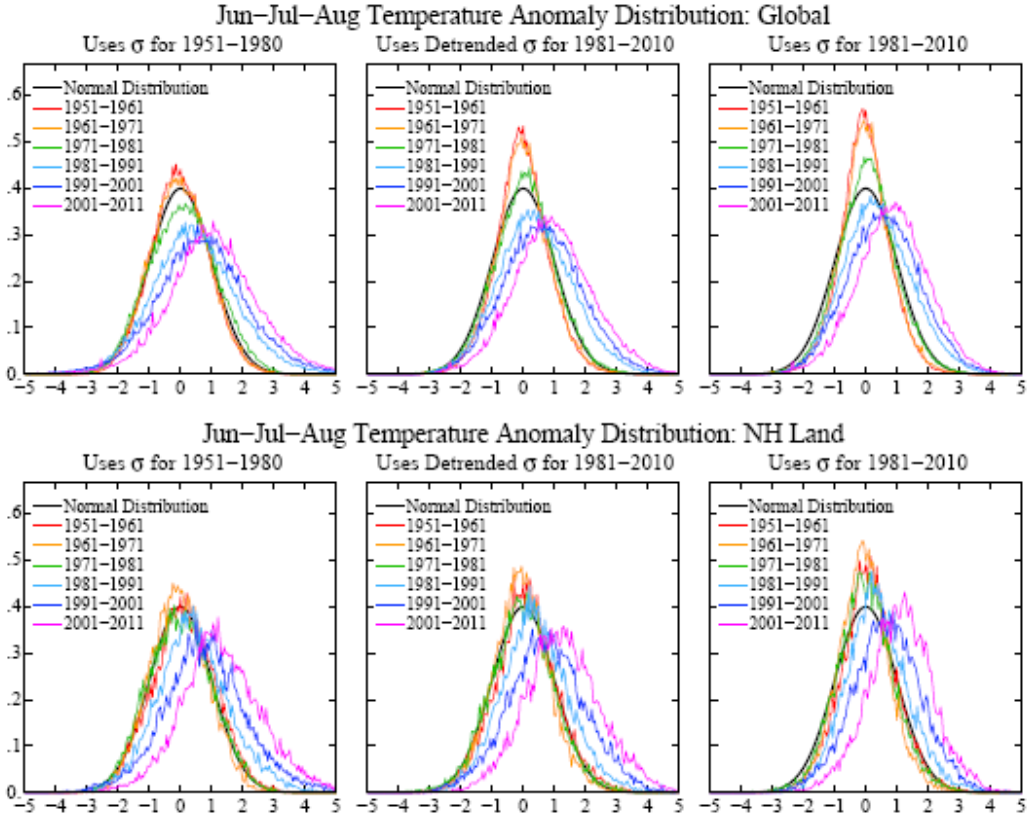

**Fig. 4.** Frequency of occurrence (y-axis) of local temperature anomalies (relative to 1951-1980 mean) divided by local standard deviation (x-axis) obtained by counting how many gridboxes have anomalies within each 0.05 interval over 11-year periods. Area under each curve is unity.

*Frequency distribution of temperature anomalies.* We first examine how the temperature anomaly distribution is changing. The Jun-Jul-Aug temperature anomaly distribution in successive decadal periods is shown in Fig. 4 for the three choices of standard deviation in Fig. 2. For comparison the normal (a.k.a. Gaussian or bell-curve) distribution of anomalies is shown by the black line. The upper row is the global result and the lower row is for Northern Hemisphere land. The data curves were obtained by binning the local anomalies divided by local standard deviation into intervals of 0.05 (i.e., by counting the number of gridboxes having a ratio within each successive 0.05 interval).

The distribution of temperature anomalies for each of the three decades within the 1951-1980 base period falls close to the normal (Gaussian) distribution with standard deviation based on that (1951-1980) period. The anomaly distributions for the three decades in 1951-1980 become more peaked than the normal distribution if the larger standard deviations of 1981-2010 are used, because of increased temperature variability in recent decades. Results for Northern Hemisphere land (lower half of Fig. 4) avoid any possible effect of an artificially small standard deviation over poorly sampled ocean areas.

The probability distribution function shifts toward the right in each successive decade in the past 30 years, the shift being somewhat larger for land areas than for the global mean. The occurrence of 3σ, 4σ and 5σ anomalies, practically absent in 1951-1980, is substantial in the past decade, consistent with the large brown areas in Fig. 3. The frequency of seasons that were cooler than the 1951-1980 average (temperature anomaly < 0°C) is obviously greatly diminished in recent decades, as we will quantify below.

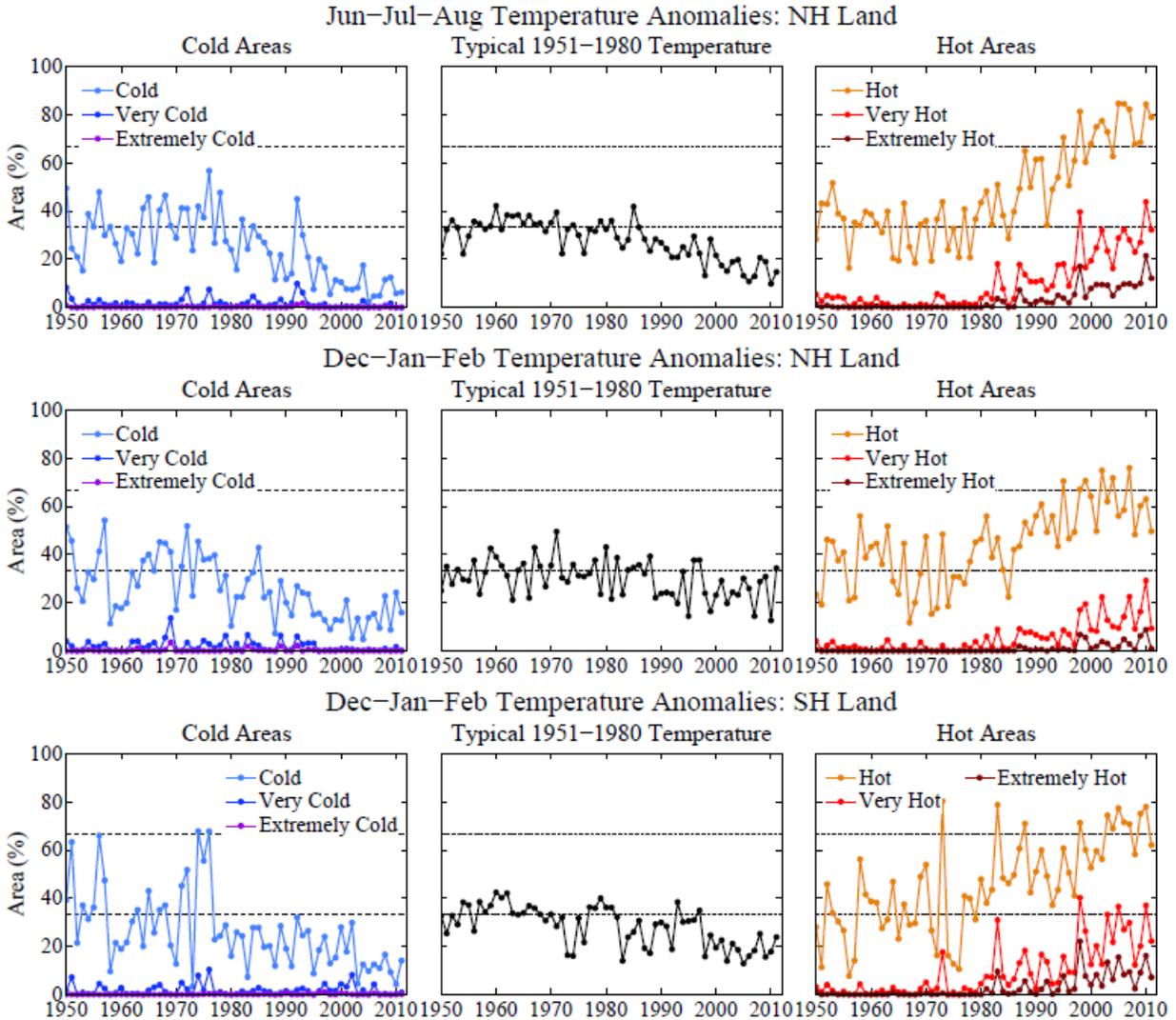

**Fig. 5.** Area covered by temperature anomalies in the categories defined as hot (> 0.43σ), very hot (> 2σ), and extremely hot (> 3σ), with analogous divisions for cold anomalies. Anomalies are relative to 1951-1980 climatology, with σ also from 1951-1980 data. Lowest row is Southern Hemisphere summer.

*Loaded climate dice.* "Loading" of the "climate dice" is one way to describe the systematic shift of the frequency distribution of temperature anomalies. Hansen et al. (3) represented the climate of 1951-1980 by colored dice with two sides colored red for "hot", two sides blue for "cold", and two sides white for near average temperatures. With a normal distribution of temperatures the dividing points are at ±0.43σ to achieve equal (one third) chances of being in each of these three categories in the period of climatology (1951-1980).

Hansen et al. (3) used a climate model to project how the odds would change due to global warming for alternative greenhouse gas scenarios. Their scenario B, which had climate forcing that turned out to be close to subsequent reality, led to four of the six dice sides being red early in the 21$^{st}$ century, based on their global climate model simulations.

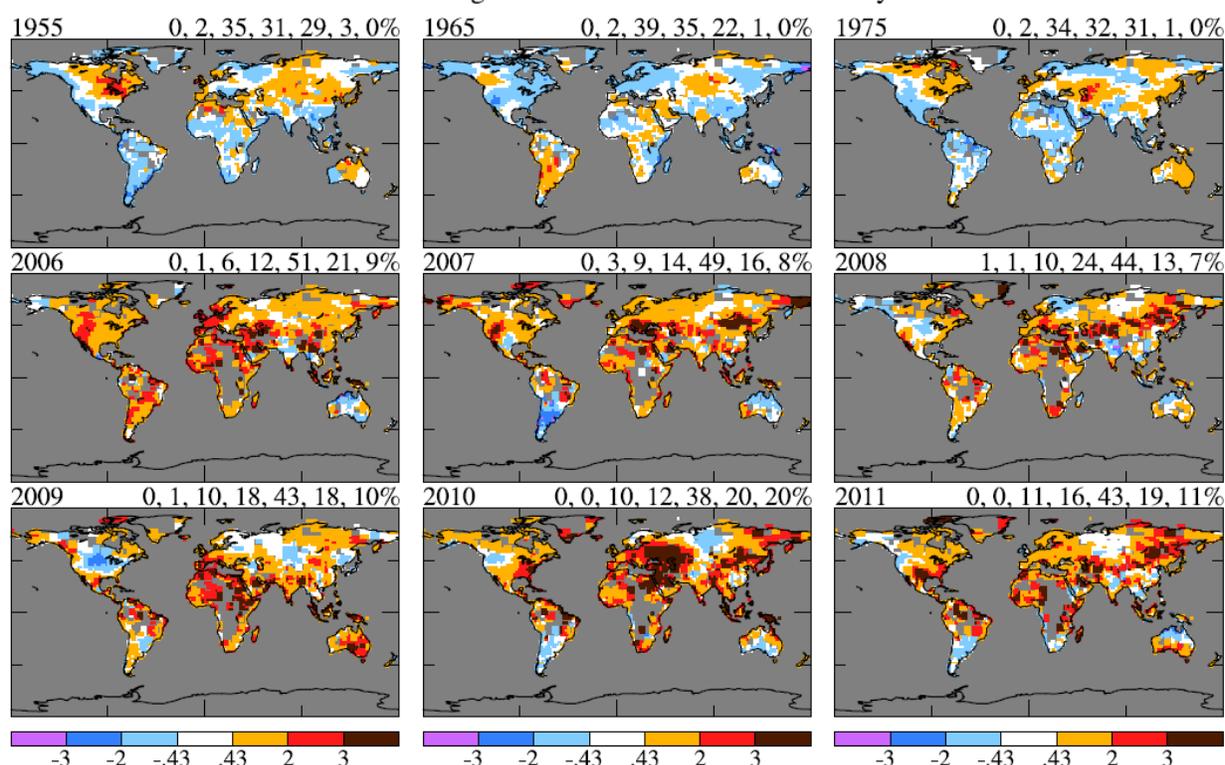

**Fig. 6.** Jun-Jul-Aug surface temperature anomalies over land in 1955, 1965, 1975 and 2006-2011 relative to 1951-1980 mean temperature in units of the local 1951-1980 standard deviation of temperature. Numbers above each map are the percent of surface area covered by each category in the color bar.

Fig. 5 reveals that the occurrence of "hot" summers (seasonal mean temperature anomaly exceeding $+0.43\sigma$) has reached the level of 67% required to make four sides of the dice red in both the Northern Hemisphere (top row) and Southern Hemisphere (bottom row). The loading of the dice in winter (middle row), i.e., the shift to unusually warm seasons, is not as great as in summer, despite the fact that observed warming in winter is larger than in summer (5). The reason for the smaller apparent change in winter is the much larger chaotic climate variability of temperature in that season, as summarized by the standard deviation (Fig. 2).

Probably the most important change is the emergence of a new category of "extremely hot" summers, more than $3\sigma$ warmer than climatology. Fig. 6 illustrates that $+3\sigma$ anomalies practically did not exist in the period of climatology (1951-1980), but in the past several years these extreme anomalies have covered of the order of 10% of the land area.

Maps analogous to Fig. 6 but for Dec-Jan-Feb are included on the web site http://www.columbia.edu/~mhs119/PerceptionsAndDice to allow examination of trends for winter and summer in both hemispheres. Winter trends in units of standard deviations are comparable to those in summer, but tend to be smaller. Warming is larger in winter than in summer, but this tends to be more than offset by the much larger natural variability in winter (Fig. 2), which makes it harder for the public to notice climate change in winter. Another factor that may affect public perception of winter warming is a tendency of the public to equate heavy snowfall and harsh winter conditions, even if temperatures are not extremely low. Observations (14, 15) confirm expectations that a warmer atmosphere holds more water vapor, and thus snowfall may increase with warming in places that remain cool enough for snow.

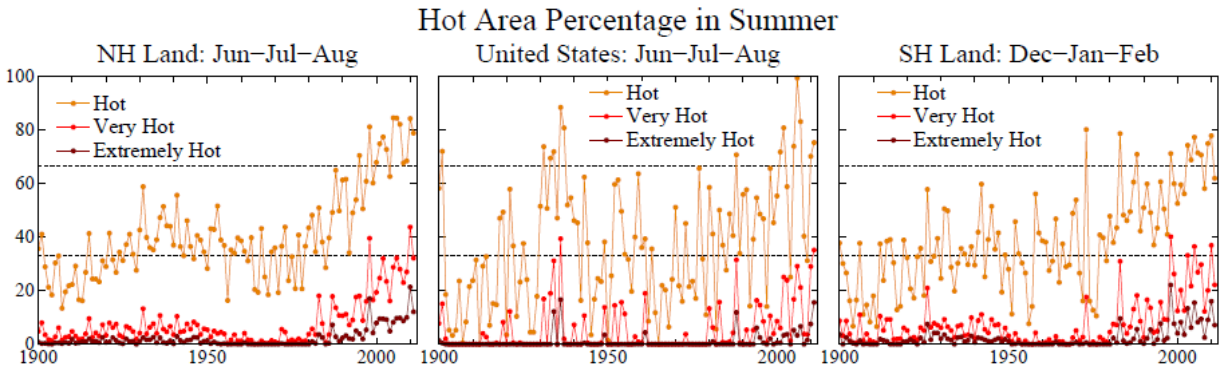

**Fig. 7.** Percent area covered by temperature anomalies in categories defined as hot (> 0.43σ), very hot (> 2σ), and extremely hot (> 3σ). Anomalies are relative to 1951-1980 climatology; σ is from 1951-80 data.

The increase, by more than a factor 10, of area covered by extreme hot summer anomalies (> +3σ ) reflects the shift of the anomaly distribution in the past 30 years of global warming, as shown succinctly in Fig. 4.  One implication of this shift is that the extreme summer climate anomalies in Texas in 2011, in Moscow in 2010, and in France in 2003 almost certainly would not have occurred in the absence of global warming with its resulting shift of the anomaly distribution.  In other words, we can say with a high degree of confidence that such extreme anomalies would not have occurred in the absence of global warming.

How will the "loading" of the climate dice continue to change in the future?  Fig. 4 provides a clear, sobering, indication.  The extreme hot tail of the distribution of temperature anomalies shifted to the right by more than +1σ in response to the global warming of about 0.5°C over the past three decades.  Additional global warming in the next 50 years, if business-as-usual fossil fuel emissions continue, is expected to be at least 1°C (4).  In that case, the further shifting of the anomaly distribution will make +3σ anomalies the norm and +5σ anomalies will be common.

*A longer time scale and regional detail.*  Jun-Jul-Aug data on a longer time scale, 1900-present, including results averaged over the 48 contiguous states of the United States, are shown in Fig. 7. The small area of the 48 states (less than 1.6% of the globe) causes temperature anomalies for the United States to be very "noisy".  Nevertheless, it is apparent that the long-term trend toward hot summers is not as pronounced in the United States as it is in hemispheric land as a whole.  Also note that the extreme summer heat of the 1930s, especially 1934 and 1936, is comparable to the most extreme recent years.

Large regional anomalies are of interest, but let us first note that the extreme anomalies of the 1930s and 1940s do not obviate the conclusion that recent global warming, with high probability, is responsible for recent extreme anomalies.  In the Supporting Information we show maps of temperature anomalies for six years with the greatest "hot" area (1931, 1934, 1936, 1941, 1947, 1953).  Those years were warmer (globally and in the United States) than the 1951-1980 mean, so it is not surprising that the area with 3σ anomalies was greater than in the 1951-1980 climatology.  The largest area of 3σ anomalies was in 1941, when it reached 2.7% of the land area.  This compares with recent values as great as 20% and an average about 10%.

Year-to-year variability, which is mainly unforced weather variability, is so large for an area the size of the United States that it is perhaps unessential to find an "explanation" for deviations from the global trend.  However, the interpretation matters, because, if the lesser warming in recent years is a statistical fluke, the United States may have in store a relatively rapid trend toward more extreme anomalies. If it is not a fluke, and if the basis for a reduced effect continues, it may continue to be difficult to garner support in the U.S. for climate mitigation.

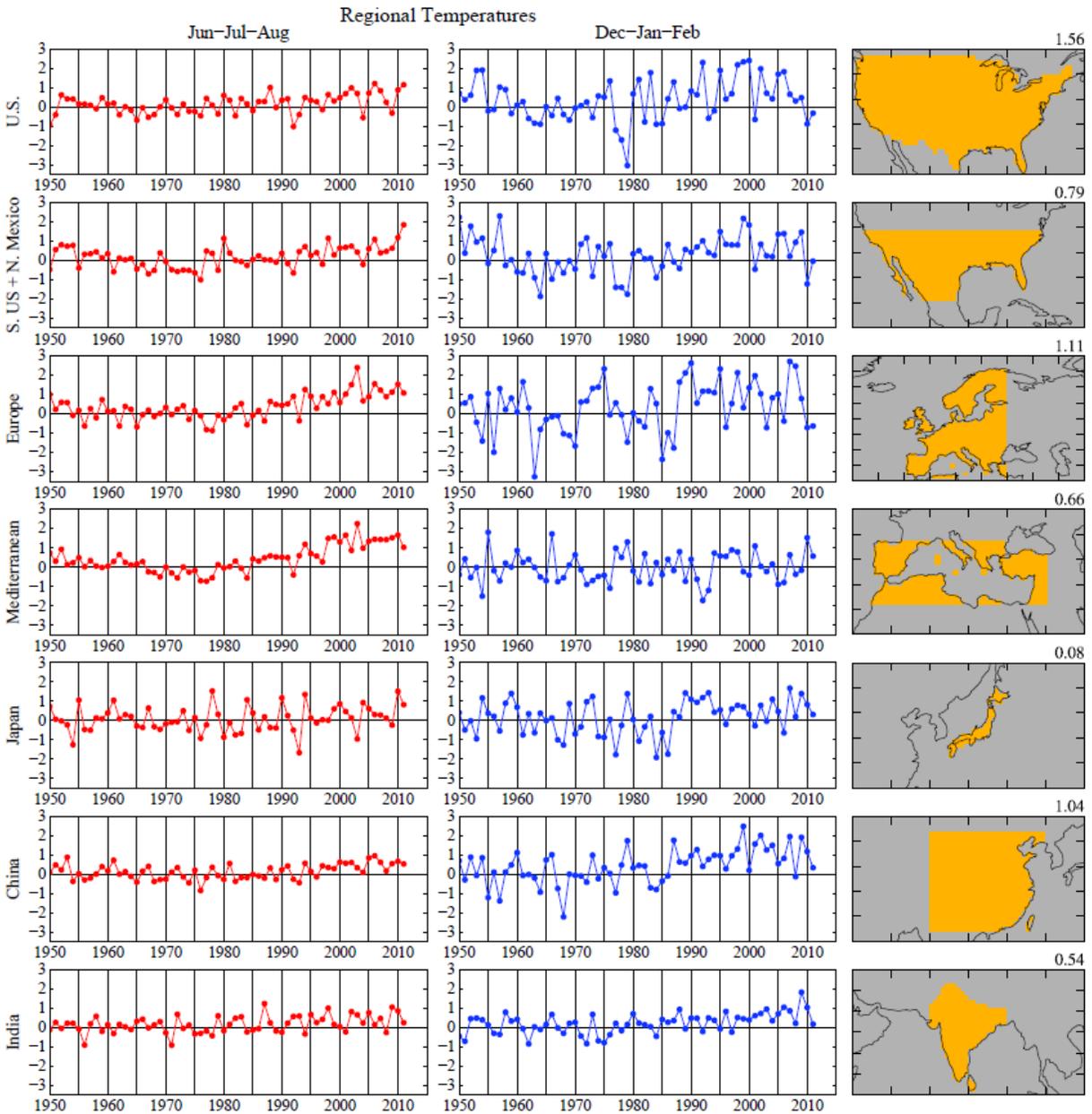

**Fig. 8.** Jun-Jul-Aug and Dec-Jan-Feb temperature anomalies (°C) for areas shown on the right.

Some researchers have suggested that the high summer temperatures and drought in the United States in the 1930s can be accounted for by sea surface temperature patterns plus natural variability (16, 17) Other researchers (18-20), have presented evidence that agricultural changes (plowing of the Great Plains) and crop failure in the 1930s contributed to changed surface albedo, aerosol (dust) production, high temperatures, and drying conditions. Also empirical evidence and climate simulations (20, 21) suggest that agricultural irrigation has a significant regional cooling effect. Such regionally-varying effects could be partly responsible for differences between observed regional temperature trends and the global trend.

Prediction of regional climate change is difficult because of the multiple factors that can affect regional climate and the high degree of chaotic (unforced) variability. In addition to a general warming trend, we might expect to find evidence in the data of a poleward shift of climatic zones. Theory and climate models indicate that the overturning tropical circulation, the Hadley cell, will expand poleward with global warming (22, 23). There is evidence in satellite and radiosonde data and reanalyses output for poleward expansion of the tropical circulation of as much as a few degrees of latitude during the past three decades (24), but changes of several of the indicators used to define the tropical boundary are not statistically significant (25).

Impacts of expansion of the overturning tropical circulation in the Northern Hemisphere might be anticipated in the southern United States and Mediterranean region in the summer, when the descending branch of the Hadley circulation extends into those areas. Despite a global increase in rainfall, regions experiencing intensification of subtropical conditions can expect periods of increased aridity and higher temperatures (26, 27), which contribute to increased forest fires that burn hotter and are more destructive (28).

We compare summer and winter temperature anomalies for several regions in Fig. 8, with the area in China being the part with most of the population. This figure reveals that even for these small regions (maximum size about 1.5% of globe) a systematic warming tendency is apparent, especially in the summer. However, at most places seasonal mean temperatures cooler than the 1951-1980 mean still occur occasionally, especially in the winter.

**Discussion**

Seasonal-mean temperature anomalies have changed dramatically in the past three decades, especially in the summer. The shift of the probability distribution (Fig. 9, left) is more than one standard deviation. In addition, the probability distribution broadens, the warming shift being greater at the high temperature tail of the distribution than at the low temperature tail.

The climate dice are now loaded to a degree that the perceptive person (old enough to remember the climate of 1951-1980) should recognize the existence of climate change. This is especially true in summer. Summers with mean temperature in the category defined as "cold" in 1951-1980 climatology (mean temperature below $-0.43\sigma$), which occurred about one-third of the time in 1951-1980, now occur with a frequency about 10%, while those in the "hot" category have increased from about 33% to about 75% (Fig. 7).

The most important change of the climate dice is the appearance of a new category of extremely hot summer anomalies, with mean temperature at least three standard deviations greater than climatology. These extreme temperatures were practically absent in the period of climatology, covering only a few tenths of one percent of the land area, but they have occurred over about 10% of land area in recent years. The increased frequency of these extreme anomalies, by more than an order of magnitude, implies that we can say with a high degree of confidence that events such as the extreme summer heat in the Moscow region in 2010 and Texas in 2011 were a consequence of global warming. Rahmstorf and Coumou (29), using a more elegant mathematical analysis, reached a similar conclusion for the Moscow anomaly.

It is not uncommon for meteorologists to reject global warming as a cause of these extreme events, offering instead a meteorological explanation. For example, it is said that the Moscow heat wave was caused by an extreme atmospheric "blocking" situation, or the Texas heat wave was caused by La Nina ocean temperature patterns. Certainly the locations of extreme anomalies in any given case depend on specific weather patterns. However, blocking patterns and La Ninas have always been common, yet the large areas of extreme warming have come into existence only with large global warming. Today's extreme anomalies occur because of simultaneous contributions of specific weather patterns and global warming.

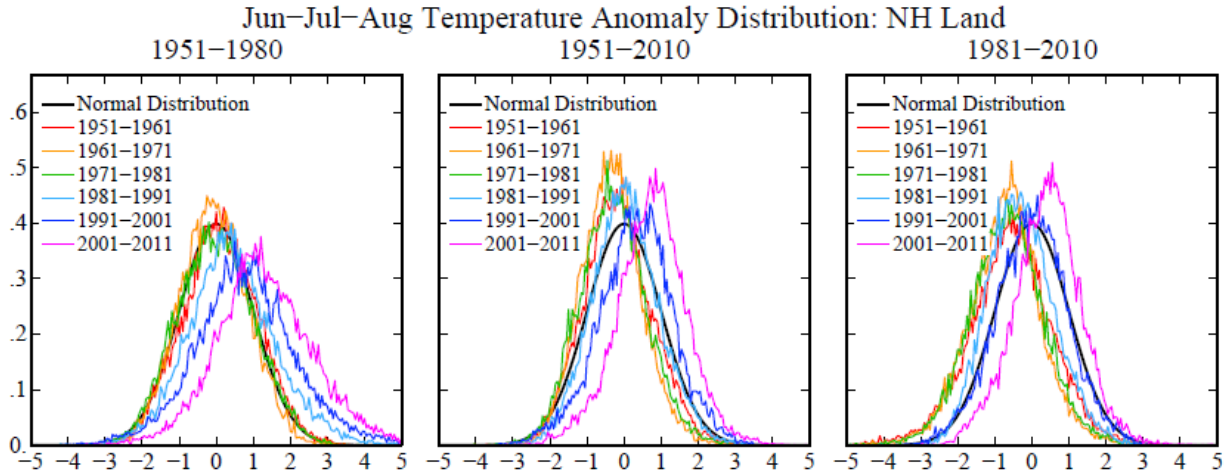

**Fig. 9.** Frequency of occurrence (y-axis) of local temperature anomalies divided by local standard deviation (x-axis) obtained by binning all local results for 11-year periods into 0.05 frequency intervals. Area under each curve is unity. Standard deviations are for the indicated base periods.

However, we must ask: do our conclusions depend on the base period chosen for climatology? Can we just redefine climatology based on the most recent decades, perhaps leading to a conclusion that the only climate change has been a small shift of mean temperature that may be insignificant?

The effect of alternative base periods on the frequency of temperature anomalies is shown in Fig. 9. Use of a recent base period alters the appearance of the probability distribution function for temperature anomalies, because the frequency of occurrence is expressed in units of the standard deviation. Because climate variability increased in recent decades, and thus the standard deviation increased, if we use the most recent decades as base period we "divide out" the increased variability. Thus the distribution function using 1981-2010 as the base period (right graph in Fig. 9) does not expose the change toward increased climate variability.

For many decades the World Meteorological Organization has used the prior three decades to define climatology (30). This is a useful procedure when the objective is to define anomalies relative to a recent period that most people will be familiar with. However, this practice tends to hide the fact that climate variability itself is changing on decadal time scales. Thus, at least for research purposes, we recommend keeping the base period fixed.

The question then becomes, what is the most appropriate base period. We argue that the appropriate base period is close to our initial choice, 1951-1980, because that was a period of relatively stable global temperature. The period 1951-1980 is also the earliest base period with good global coverage of meteorological stations, including Antarctica.

Another merit of 1951-1980 is that it is more representative of the Holocene (31) than any later period would be. This is important because we would prefer a base period to which plant and animal life on the planet are adapted. Hansen and Sato (7) argue that the climate of the most recent few decades is probably warmer than prior Holocene levels, based on the fact that the major ice sheets in both hemispheres are presently losing mass rapidly (9) and global sea level is rising at a rate of more than 3 m/millennium (32), which is much greater than the slow rate of sea level change (less than 1 m/millennium) in the latter half of the Holocene (33).

Changes of global temperature are likely to have their greatest practical impact via effects on the hydrologic cycle. The +3σ summer anomalies are usually in places experiencing an extended period of high atmospheric pressure. With the amplification of global warming and ubiquitous surface heating from elevated greenhouse gas amounts, extreme drought conditions can develop.

However, the other extreme of the hydrologic cycle, unusually heavy rainfall and floods, is also amplified by global warming. A warmer world is expected to have more extreme rainfall occurrences because the amount of water vapor that the atmosphere holds increases rapidly with temperature, a tendency confirmed by observations. Indeed, rainfall data reveal significant increases of heavy precipitation over much of Northern Hemisphere land and in the tropics (23) and attribution studies link this intensification of rainfall and floods to human-made global warming (34-36).

Extreme heat waves and record floods receive most public attention, yet we wonder if there is not also a more pervasive effect of warming that affects almost everyone. Natural ecosystems are adapted to the stable climate of the Holocene. Although climate fluctuations are normal, the rapid global trend of the past three decades, from an already warm level, is highly unusual. The fact that warmer winters have led to an epidemic of pine bark beetles and widespread destruction of forests in Canada and western United States (34) is well known. However, there are surely more pervasive effects of this strong warming trend. Climate change of recent decades is already affecting the geographical and seasonal range of animals, birds and insects (37) to a degree that is sometimes noticeable to the public (38). These changes should be more and more perceptible to the public during the next decade as the frequency distribution of temperature anomalies continues to shift toward higher values.

Although species migrate to stay within climate zones in which they can survive, continued climate shift at the rate of the past three decades could take an enormous toll on planetary life. It is estimated that 21-52% of the species on Earth will be committed to extinction, if global warming approaches 3°C by the end of the century (23). Fortunately, climate scenarios are also conceivable in which such large warming is avoided by placing a rising price on carbon emissions, thus moving the world to a clean energy future fast enough to limit peak global warming to several tenths of a degree Celsius above the current level (39). It is argued (39) that such a scenario is needed if we are to preserve life on Earth as we know it.

**Acknowledgments.** We thank Tom Karl and Andrew Weaver for helpful reviews that significantly improved the paper.

**References**

1. Rabe, B.G. and Borick, C.P. (2012) Fall 2011 national survey of American public opinion on climate change, Issues in Governance Studies. (Brookings Institution, Washington, D.C.).
2. Hansen, J., *et al.*, 2008: Target Atmospheric $CO_2$: Where Should Humanity Aim? *The Open Atmos Sci J,* **2**, 217-231.
3. Hansen, J., *et al.*, 1988: Global climate changes as forecast by Goddard Institute for Space Studies 3-dimensional model. *J Geophys Res - Atmos,* **93**, 9341-9364.

4. Intergovernmental Panel on Climate Change (IPCC), 2007: *Climate Change 2007: The Physical Science Basis*, Solomon, S.*, et al.* eds., Cambridge University Press, 996 pp.
5. Hansen, J., Ruedy, R., Sato, M., and Lo, K., 2010: Global surface temperature change. *Rev Geophys,* **48** RG4004.
6. Menne, M.J., Williams, C.N., and Vose, R.S., 2009: The US Historical Climatology Network Monthly Temperature Data, Version 2. *Bull Am Met Soc,* **90**, 993-1007.
7. Hansen, J.E. and Sato, M. (2012) Paleoclimate implications for human-made climate change *Climate Change: Inferences from Paleoclimate and Regional Aspects*, eds Berger, A., Mesinger, F., and Sijacki, D., Springer, 244 pp.
8. Velicogna, I., 2009: Increasing rates of ice mass loss from the Greenland and Antarctic ice sheets revealed by GRACE. *Geophys Res Lett,* **36** L19503.
9. Rignot, E., Velicogna, I., van den Broeke, M.R., Monaghan, A., and Lenaerts, J., 2011: Acceleration of the contribution of the Greenland and Antarctic ice sheets to sea level rise. *Geophys Res Lett,* **38** L05503.
10. Nerem, R.S., Leuliette, E., and Cazenave, A., 2006: Present-day sea-level change: A review. *Cr Geosci,* **338**, 1077-1083.
11. Jones, P.D., New, M., Parker, D.E., Martin, S., and Rigor, I.G., 1999: Surface air temperature and its changes over the past 150 years. *Rev Geophys,* **37**, 173-199.
12. Brohan, P., Kennedy, J.J., Harris, I., Tett, S.F.B., and Jones, P.D., 2006: Uncertainty estimates in regional and global observed temperature changes: A new data set from 1850. *J Geophys Res,* **111** D12106.
13. Smith, T.M., Reynolds, R.W., Peterson, T.C., and Lawrimore, J., 2008: Improvements to NOAA's historical merged land-ocean surface temperature analysis (1880-2006). *J Clim,* **21**, 2283-2296.
14. Reynolds, R.W., Zhang, H.M., Smith, T.M., Gentemann, C.L., and Wentz, F., 2005: Impacts of in situ and additional satellite data on the accuracy of a sea-surface temperature analysis for climate. *Int J Climatol,* **25**, 857-864.
15. Yu, L.S. and Weller, R.A., 2007: Objectively analyzed air-sea heat fluxes for the global ice-free oceans (1981-2005). *Bull Am Met Soc,* **88**, 527.
16. Nigam, S., Guan, B., and Ruiz-Barradas, A., 2011: Key role of the Atlantic Multidecadal Oscillation in 20th century drought and wet periods over the Great Plains. *Geophys Res Lett,* **38** L16713.
17. Hoerling, M.*, et al.*, 2012: On the increased frequency of Mediterranean drought. *J Clim (in press)*.
18. Cook, B.I., Miller, R.L., and Seager, R., 2009: Amplification of the North American "Dust Bowl" drought through human-induced land degradation. *Proc Natl Acad Sci USA,* **106**, 4997-5001.
19. Cook, B.I., Terando, A., and Steiner, A., 2010: Ecological forecasting under climatic data uncertainty: a case study in phenological modeling. *Environ Res Lett,* **5**, 044014.
20. Cook, B.I., Puma, M.J., and Krakauer, N.Y., 2011: Irrigation induced surface cooling in the context of modern and increased greenhouse gas forcing. *Clim Dyn,* **37**, 1587-1600.
21. Puma, M.J. and Cook, B.I., 2010: Effects of irrigation on global climate during the 20th century. *J Geophys Res,* **115** D16120.
22. Held, I.M. and Soden, B.J., 2006: Robust responses of the hydrological cycle to global warming. *J. Clim.,* **19**, 5686-5699.

23. Intergovernmental Panel on Climate Change (IPCC), 2007: *Climate Change 2007, Impacts, Adaptation and Vulnerability*, Parry, M.L., Canziani, O.F., Palutikof, J.P., Van Der Linden, P.J., and Hanson, C.E. eds., Cambridge Univ Press, 996 pp.
24. Seidel, D.J., Fu, Q., Randel, W.J., and Reichler, T.J., 2008: Widening of the tropical belt in a changing climate. *Nat Geosci,* **1**, 21-24.
25. Davis, S.M. and Rosenlof, K.H., 2012 A multi-diagnostic intercomparison of tropical width time series using reanalyses and satellite observations. *J Clim,* **25**, 1061-1078.
26. Barnett, T.P.*, et al.*, 2008: Human-induced changes in the hydrology of the western United States. *Science,* **319**, 1080-1083.
27. Levi, B.G., 2008: Trends in the hydrology of the western US bear the imprint of manmade climate change. *Phys Today,* **61**, 16-18.
28. Westerling, A.L., Hidalgo, H.G., Cayan, D.R., and Swetnam, T.W., 2006: Warming and earlier spring increase western US forest wildfire activity. *Science,* **313**, 940-943.
29. Rahmstorf, S. and Coumou, D., 2011: Increase of extreme events in a warming world. *Proc Natl Acad Sci USA,* **108**, 17905-17909.
30. Arguez, A. and Vose, R.S., 2011: The Definition of the Standard WMO Climate Normal The Key to Deriving Alternative Climate Normals. *B Am Meteorol Soc,* **92**, 699-U345.
31. Mayewski, P.A.*, et al.*, 2004: Holocene climate variability. *Quaternary Res,* **62**, 243-255.
32. Church, J.A. and White, N.J., 2011: Sea-Level Rise from the Late 19th to the Early 21st Century. *Surv Geophys,* **32**, 585-602.
33. Lambeck, K. and Chappell, J., 2001: Sea level change through the last glacial cycle. *Science,* **292**, 679-686.
34. Raffa, K.F.*, et al.*, 2008: Cross-scale drivers of natural disturbances prone in anthropogenic amplification: The dynamics of bark beetle eruptions. *BioSci,* **58**, 501-517.
35. Pall, P.*, et al.*, 2011: Anthropogenic greenhouse gas contribution to flood risk in England and Wales in autumn 2000. *Nature,* **470**, 382-385.
36. Min, S.K., Zhang, X.B., Zwiers, F.W., and Hegerl, G.C., 2011: Human contribution to more-intense precipitation extremes. *Nature,* **470**, 378-381.
37. Parmesan, C., 2006: Ecological and evolutionary responses to recent climate change. *Ann Rev Ecol Evol S,* **37**, 637-669.
38. Hansen, J. (2008) Tipping point: Perspective of a climatologist. *The State of the Wild: A global Portrait of Wildlife, Wild Lands, and Oceans*, ed Fearn, E., Wildlife Conservation Society/Island Press, 6-15 pp.
39. Hansen, J.*, et al.* (2012) Scientific case for avoiding dangerous climate change to protect young people and nature (Cornell University Library, arXiv.org).

**Supporting Information**

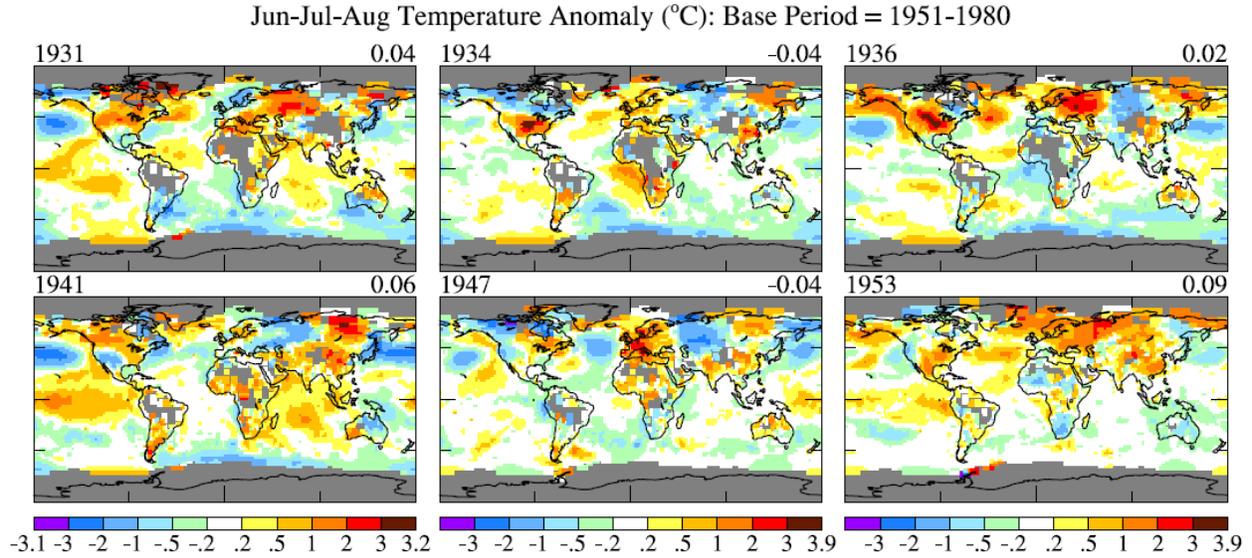

**Fig. S1.** Jun-Jul-Aug surface temperature anomalies in 1931, 1934, 1936, 1941, 1947, 1953. Number on upper right is the global mean (average over all area with data).

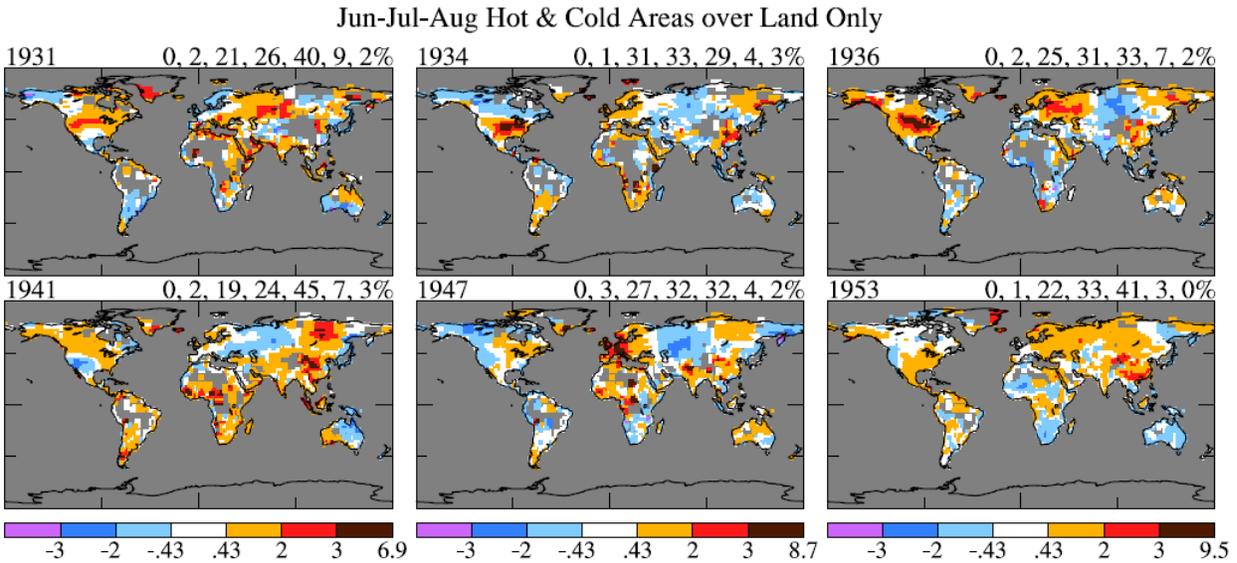

**Fig. S2.** Jun-Jul-Aug surface temperature anomalies over land in 1931, 1934, 1936, 1941, 1947, 1953 relative to 1951-1980 mean temperature in units of the local 1951-1980 standard deviation of temperature. Numbers above each map are the percent of surface area covered by each category in the color bar.

19